\documentclass[aps,prl,preprint,tightenlines,superscriptaddress,showpacs,byrevtex]{revtex4}

\usepackage{graphicx}
\usepackage{xspace}
\usepackage{epsfig}
\usepackage{multirow}

\newcommand{\acp}{\ensuremath{\mathcal{A}_{CP}}\xspace}

\def\myspecial#1{}                   
\def\calL{{\mathcal L}}

\begin{document}

\myspecial{!userdict begin /bop-hook{gsave 300 50 translate 5 rotate
    /Times-Roman findfont 18 scalefont setfont
    0 0 moveto 0.70 setgray
    (\mySpecialText)
    show grestore}def end}


\vspace*{-3\baselineskip}
\resizebox{!}{3cm}{\includegraphics{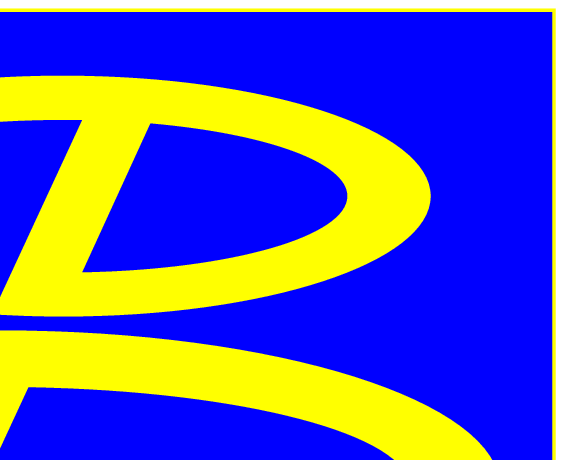}}

\preprint{\vbox{ \hbox{   }
                 \hbox{   }
                 \hbox{   }
                 \hbox{Belle Prerpint 2004-20}
                 \hbox{KEK Prerpint 2004-28}
}}

\title{\quad\\[0.5cm] \Large
Improved Measurements of Partial Rate Asymmetry in $B \to h h$ Decays}

\tighten
\affiliation{Budker Institute of Nuclear Physics, Novosibirsk}
\affiliation{Chiba University, Chiba}
\affiliation{Chonnam National University, Kwangju}
\affiliation{University of Cincinnati, Cincinnati, Ohio 45221}
\affiliation{Gyeongsang National University, Chinju}
\affiliation{University of Hawaii, Honolulu, Hawaii 96822}
\affiliation{High Energy Accelerator Research Organization (KEK), Tsukuba}
\affiliation{Hiroshima Institute of Technology, Hiroshima}
\affiliation{Institute of High Energy Physics, Chinese Academy of Sciences, Beijing}
\affiliation{Institute of High Energy Physics, Vienna}
\affiliation{Institute for Theoretical and Experimental Physics, Moscow}
\affiliation{J. Stefan Institute, Ljubljana}
\affiliation{Kanagawa University, Yokohama}
\affiliation{Korea University, Seoul}
\affiliation{Kyoto University, Kyoto}
\affiliation{Kyungpook National University, Taegu}
\affiliation{Swiss Federal Institute of Technology of Lausanne, EPFL, Lausanne}
\affiliation{University of Ljubljana, Ljubljana}
\affiliation{University of Maribor, Maribor}
\affiliation{University of Melbourne, Victoria}
\affiliation{Nagoya University, Nagoya}
\affiliation{Nara Women's University, Nara}
\affiliation{National Kaohsiung Normal University, Kaohsiung}
\affiliation{National United University, Miao Li}
\affiliation{Department of Physics, National Taiwan University, Taipei}
\affiliation{H. Niewodniczanski Institute of Nuclear Physics, Krakow}
\affiliation{Nihon Dental College, Niigata}
\affiliation{Niigata University, Niigata}
\affiliation{Osaka City University, Osaka}
\affiliation{Osaka University, Osaka}
\affiliation{Panjab University, Chandigarh}
\affiliation{Peking University, Beijing}
\affiliation{Saga University, Saga}
\affiliation{University of Science and Technology of China, Hefei}
\affiliation{Seoul National University, Seoul}
\affiliation{Sungkyunkwan University, Suwon}
\affiliation{University of Sydney, Sydney NSW}
\affiliation{Tata Institute of Fundamental Research, Bombay}
\affiliation{Toho University, Funabashi}
\affiliation{Tohoku Gakuin University, Tagajo}
\affiliation{Tohoku University, Sendai}
\affiliation{Department of Physics, University of Tokyo, Tokyo}
\affiliation{Tokyo Institute of Technology, Tokyo}
\affiliation{Tokyo Metropolitan University, Tokyo}
\affiliation{Tokyo University of Agriculture and Technology, Tokyo}
\affiliation{University of Tsukuba, Tsukuba}
\affiliation{Virginia Polytechnic Institute and State University, Blacksburg, Virginia 24061}
\affiliation{Yonsei University, Seoul}
  \author{Y.~Chao}\affiliation{Department of Physics, National Taiwan University, Taipei} 
  \author{P.~Chang}\affiliation{Department of Physics, National Taiwan University, Taipei} 
  \author{K.~Abe}\affiliation{High Energy Accelerator Research Organization (KEK), Tsukuba} 
  \author{K.~Abe}\affiliation{Tohoku Gakuin University, Tagajo} 
  \author{H.~Aihara}\affiliation{Department of Physics, University of Tokyo, Tokyo} 
  \author{Y.~Asano}\affiliation{University of Tsukuba, Tsukuba} 
  \author{T.~Aushev}\affiliation{Institute for Theoretical and Experimental Physics, Moscow} 
  \author{S.~Bahinipati}\affiliation{University of Cincinnati, Cincinnati, Ohio 45221} 
  \author{A.~M.~Bakich}\affiliation{University of Sydney, Sydney NSW} 
  \author{Y.~Ban}\affiliation{Peking University, Beijing} 
% ...
\collaboration{The Belle Collaboration}

\begin{abstract}
We report improved measurements of the partial rate asymmetry (\acp) in
$B \to h h$ decays with 140~fb$^{-1}$ of data collected with the Belle
detector at the KEKB $e^+e^-$ collider. Here $h$ stands for
a charged or neutral pion or kaon and in total five decay modes are 
included: $K^\mp\pi^\pm, K^0_S \pi^\mp, K^\mp \pi^0, \pi^\mp\pi^0$ and 
$K^0_S \pi^0$. The flavor of the last decay mode is determined from the 
accompanying $B$ meson. Using a data sample 4.7 times larger than that of
our previous measurement, we find $\acp(K^\mp\pi^\pm)$
$-0.088\pm0.035\pm0.013$, $2.4 \sigma$ from zero.
Results for other decay modes are also presented.

\end{abstract}

\maketitle

{\renewcommand{\thefootnote}{\fnsymbol{footnote}}

\setcounter{footnote}{0}

\normalsize

\newpage
In the Standard Model (SM) $CP$ violation arises via the interference of at
least two diagrams with comparable amplitudes but different $CP$ conserving
and violating phases. Two-body charmless hadronic $B$ decays may thus
be used to access $CP$ violation because of the possibility of large
interference between tree and penguin diagrams. The angle $\phi_1$ of the
unitarity triangle has been measured with good accuracy using $b\to c\bar{c} s$
transitions \cite{2phi1,2beta}. The focus of the $B$ factory experiments has
now shifted to measurement of the other two angles, $\phi_2$ and $\phi_3$,
and to the search for direct $CP$ violation.

Recent theoretical work suggests that $\phi_2$ and $\phi_3$ could be
constrained from measurements of branching fractions and partial rate
asymmetries in $B \to hh$ decays. However, different theoretical approaches
such as perturbative QCD (PQCD) \cite{0306004} and QCD factorization (QCDF)
\cite{0307095,013004,0104110} lead to different predictions for the
partial rate $CP$ asymmetry, defined as
\begin{eqnarray}
\acp=\frac{N(\overline B \to \overline f)-N(B \to f)}
{N(\overline B \to \overline f)+N(B \to f)},
\end{eqnarray} 
where $N(\overline B \to \overline f)$ is the yield for the
$\overline B \to h h$ decay and $N(B \to f)$ denotes that of the
charge-conjugate mode.
For instance, QCDF predicts a
positive asymmetry of less than 10\% in $B\to K\pi$ while PQCD favors a
negative value ranging from $-10\%$ to $-30\%$. Although there are large
uncertainties related to hadronic effects, more precise measurements of
partial rate asymmetries would discriminate between the theoretical
approaches. Furthermore, if a sizable asymmetry is observed for a mode
where only one diagram contributes in the SM, this may be an indication of
new physics.

In the search reported in this paper, five decay modes are 
considered: $K^\mp\pi^\pm, K^0_S\pi^\mp, K^\mp\pi^0, \pi^\mp\pi^0$ and
$K^0_S\pi^0$. (In the following, charge-conjugate modes are implied in the
following unless otherwise stated.) The data sample used in this study
corresponds to
140 fb$^{-1}$ ($152 \times 10^6~B\overline B$ pairs), which is 
1.8 times larger than that used in the previous study for the $K^0_S\pi^{-}$ 
mode \cite{kspi}, and is 4.7 times larger for the other modes \cite{hh}.
The observation of large $CP$ violation in $B^0\to \pi^+\pi^-$
at Belle has been recently reported elsewhere \cite{apipi}.

The data were collected with the Belle detector at the KEKB $e^+ e^-$
collider. KEKB is an asymmetric collider (3.5 on 8 GeV) that operates
at the $\Upsilon(4S)$ resonance ($\sqrt{s}=10.58 ~{\rm GeV}$).
The Belle detector is a large-solid-angle magnetic spectrometer consisting
of a three-layer silicon vertex detector, a 50-layer central drift chamber
(CDC), a system of aerogel threshold Cherenkov counter (ACC),
time-of-flight scintillation counters (TOF), and an array of CsI
crystals (ECL) located inside a superconducting solenoid that provides
1.5 T magnetic field. An iron flux return located outside is
instrumented as a $K_L^0$ and muon identifier. A detailed description of the
Belle detector can be found in Ref.~\cite{belle}.

The $B$ candidate selection is the same as described in Ref.~\cite{btohh}.
Charged tracks are required to originate from the interaction point (IP).
Charged kaons and pions are identified using $dE/dx$
information and Cherenkov light yields in the ACC.
The $dE/dx$ and ACC information  are combined to form
a $K$-$\pi$ likelihood ratio, 
$\mathcal{R}(K\pi) = \mathcal{L}_K/(\mathcal{L}_K+\mathcal{L}_\pi)$,
where $\mathcal{L}_{K/\pi}$
is the likelihood of kaon/pion. Charged tracks with $\mathcal{R}(K\pi)>0.6$ are
regarded as kaons and $\mathcal{R}(K\pi)<0.4$ as pions.
The $K/\pi$ identification efficiencies and misidentification
rates are determined from a sample of kinematically identified
$D^{*+}\to D^0\pi^+, D^0\to K^-\pi^+$ decays. For positively charged tracks,
the identification efficiencies and fake rates for $K(\pi)$ are
$83.7\pm0.2\%~(91.3\pm0.2\%)$ and $10.7\pm0.2\%~(5.1\pm0.1\%)$ and for
negatively charged tracks, they are $84.7\pm0.2\%~(90.5\pm0.2\%)$ and
$10.0\pm0.2\%~(5.7\pm0.1\%)$, respectively. Furthermore,
charged tracks that are positively identified as electrons are rejected.      

Candidate $\pi^0$ mesons are reconstructed by combining two photons with
invariant mass between 115 MeV/$c^2$ and 152 MeV/$c^2$, which corresponds to 
$\pm2.5$ standard deviations. Each photon is required to have a minimum
energy of 50 MeV in the barrel region ($32^\circ < \theta_\gamma < 129^\circ$)
or 100 MeV in the end-cap region ($17^\circ < \theta_\gamma < 32^\circ$ or
$129^\circ < \theta_\gamma < 150^\circ$), where $\theta_\gamma$ denotes the
polar angle of the photon with respect to the beam line.
To further reduce the combinatorial background, $\pi^0$ candidates 
with small decay angles ($\cos\theta^* >0.95$) are rejected, where
$\theta^*$ is the angle between a $\pi^0$ boost direction from the
laboratory frame and its $\gamma$ daughters in the $\pi^0$ rest frame.
Candidate $K^0_S$ mesons are reconstructed from pairs of oppositely charged
tracks with invariant mass in the range
$480~{\rm MeV}/c^2 < M_{\pi\pi} < 516~{\rm MeV}/c^2$.
Each candidate must have a displaced vertex and a flight direction consistent
with a $K^0_S$ originating from the IP.

Two variables are used to identify $B$ candidates: the beam-constrained mass,
$M_{\rm bc} =  
\sqrt{E^{*2}_{\mbox{\scriptsize beam}} - p_B^{*2}}$, and the energy difference,
$\Delta E = E_B^* - E^*_{\mbox{\scriptsize beam}}$, where 
$E^*_{\mbox{\scriptsize beam}}$ is the beam energy and $E^*_B$ and $p^*_B$ are
the reconstructed energy and momentum of the $B$ candidates in the
center-of-mass (CM) frame. Events with 
$M_{\rm bc} > 5.20$ GeV/$c^2$ and $-0.3~{\rm GeV} < \Delta E < 0.3~{\rm GeV}$
are selected for the final analysis for modes with $\pi^0$'s in the final state
and $-0.3~{\rm GeV} < \Delta E < 0.5~{\rm GeV}$ for the modes without. 
For the $K^0_S\pi^0$ mode, events with $\Delta E <-0.1$ GeV are excluded to
remove the background from charmless $B$ decays. 

The dominant background is from $e^+e^- \to q\bar q ~( q=u,d,s,c )$ continuum
events. To distinguish the signal from the jet-like continuum background,
event topology variables and $B$ flavor tagging information are employed.
We combine a set of modified Fox-Wolfram moments \cite{pi0pi0} into a
Fisher discriminant. The probability density functions (PDF) for this
discriminant, and the cosine of the angle between the $B$ flight
direction and the $z$ axis, are obtained using signal and continuum
Monte Carlo (MC) events. These two variables are then combined to form
a likelihood ratio
$\mathcal{R} = {\calL}_s/({\calL}_s + {\calL}_{q \bar{q}})$,
where ${\calL}_{s (q \bar{q})}$ is
the product of signal ($q \bar{q}$) probability densities. Additional
background discrimination is provided by $B$ flavor tagging.
The Belle standard flavor tagging algorithm \cite{tagging} gives two
outputs: a discrete variable indicating the flavour of the tagging $B$
and the MC-determined dilution factor $r$,
which ranges from zero for no flavor information to unity for unambiguous
flavor assignment. An event that contains
a lepton ($r$ close to unity) is more likely to be a $B \overline B$ event
so a looser $\mathcal{R}$ requirement can be applied. We combine $\mathcal{R}$
and $r$ to form a multi-dimensional likelihood ratio (MDLR), defined as
${\calL}_s^{\rm{MDLR}}/
({\calL}_s^{\rm{MDLR}} + {\calL}_{q\bar{q}}^{\rm{MDLR}})$, where
${\calL}^{\rm{MDLR}}_{s(q\bar{q})}$ denotes the likelihood of the signal
($q\bar{q}$) in  the $\mathcal{R}$--$r$ space. The continuum background is
reduced by applying a selection requirement on the MDLR. The requirement for
each mode is optimized according to the figure of merit defined as
$N_s^{exp}/\sqrt{N_s^{exp}+N_{q\bar{q}}^{exp}}$, where $N_s^{exp}$ denotes
the expected signal yields based on our previous branching fraction
measurements \cite{btohh} and $N_{q\bar{q}}^{exp}$ denotes the expected
$q\bar q$ yields from sideband data ($M_{\rm bc}<5.26$ GeV/$c^2$).
A typical requirement suppresses 94--98\% of the continuum background while
retaining 50--65\% of the signal.

Backgrounds from $\Upsilon(4S) \to B\overline B$ events are investigated using
a large MC sample. After the MDLR requirement, we find a small charmless
three-body background at low $\Delta E$, and reflections
from other $B\to hh$ decays due to $K$-$\pi$ misidentification.
   
The signal yields are extracted by applying unbinned two dimensional
maximum likelihood (ML) fits to the ($M_{\rm bc}$ and $\Delta E$)
distributions of the $B$ and $\overline B$ samples.
The likelihood for each mode is defined as
\begin{eqnarray}
\mathcal{L} & = &\frac{e^{-\sum_j N_j}}{N!} 
\times \prod_i (\sum_j N_j \mathcal{P}_j) \;\;\; \mbox{and} \\ 
\mathcal{P}_j & = &\frac{1}{2}[1- q_i \cdot \acp{}_j ]
P_j(M_{{\rm bc}i}, \Delta E_i),  
\end{eqnarray}
where $N$ is the total number of events, $i$ is the identifier of
the $i$-th event, $P(M_{\rm bc}, \Delta E)$ is the two-dimensional PDF of
$M_{\rm bc}$ and $\Delta E$, $q$ indicates the $B$ meson flavor,
$B (q=+1)$ or $\overline{B} (q=-1)$, $N_j$ is the number of events for the
category $j$, which corresponds to either signal, $q\bar{q}$ continuum,
a reflection due to $K$-$\pi$ misidentification, or
background from other charmless three-body $B$ decays.  

Unlike the other four $hh$ decay modes, the flavor of the $B$ meson in 
the $K^0_S\pi^0$ channel is not self-tagged and must be determined from
the accompanying $B$ meson. To account for the effect of $B\overline{B}$
mixing and imperfect tagging, the
term $\acp$ for the signal in Eq. 3 has to be replaced by
$\acp (1-2\chi_d)(1-2 w_l)$, where $\chi_d=0.181\pm0.004$ \cite{PDG}
is the time-integrated mixing parameter and and $w_l$ is the
wrong-tag fraction. The $K^0_S\pi^0$ sample is divided into six $r$-bins,
and the $r$-dependent wrong-tag fractions, $w_l$ ($l=1,\ldots,6$),
are determined using a high statistics sample of self-tagged
$B^0 \to D^{(*)-} \pi^+, D^{*-} \rho^+$ and $D^{*-} \ell^+ \nu$ 
events and their charge conjugates \cite{w_r}. No $K$-$\pi$ misidentification
reflection or other backgrounds from charmless three-body $B$ decays are
included in this mode.

The yields and asymmetries for the signal and backgrounds
are allowed to float in all modes except for $K^0_S \pi^0$, in which
the background asymmetry is set to zero. Since the $K^+\pi^0$ and
$\pi^+\pi^0$ reflections are difficult to distinguish with $\Delta E$
and $M_{\rm bc}$, we fit these two modes simultaneously with a fixed
reflection-to-signal ratio based on the $K$-$\pi$ identification efficiencies
and fake rates. All the signal PDFs ($P(M_{\rm bc},\Delta E)$) are obtained
using MC simulation. No strong correlations between 
$M_{\rm bc}$ and $\Delta E$ are found for the $K^0_S\pi^-$ and $K^-\pi^+$
signals. Therefore, their PDFs are modeled by products of
single Gaussians for $M_{\rm bc}$ and double Gaussians for $\Delta E$.
For the modes with $\pi^0$'s in the final state, there are correlations
between $M_{\rm bc}$ and $\Delta E$ in the tails of the signals; hence,
their PDFs are described by smoothed two-dimensional histograms.
Discrepancies between the signal peak positions in data and MC are calibrated
using $B^+ \to \overline{D}{}^0\pi^+$ decays, where the
$\overline{D}{}^0 \to K^+\pi^-\pi^0$ sub-decay is used for the modes with a
$\pi^0$ meson while $\overline{D}{}^0\to K^+\pi^-$ is used for the other modes.
The MC-predicted $\Delta E$ resolutions are verified using the
invariant mass distributions of high momentum $D$ mesons. The decay mode 
$\overline{D}{}^0\to K^+\pi^-$ is used for $B^0\to K^+ \pi^-$, 
$D^+\to K^0_S\pi^+$ for $B^+\to K^0_S \pi^+$ and
$\overline{D}{}^0\to K^+\pi^-\pi^0$ for the modes with a $\pi^0$ in the final
state. The parameters that describe the shape of the PDFs are fixed in all of
the fits.

The background PDFs for continuum events are modeled by an ARGUS function
\cite{argus} for the $M_{\rm bc}$ distribution and a 1st or 2nd order
polynomial for the $\Delta E$ distribution, assuming no correlations between
the two. A large MC sample is used to investigate the background 
from charmless $B$ decays and a smoothed two-dimensional histogram is taken
as the PDF.  

\begin{figure*}
\hspace{-1.0cm}
\epsfig{file=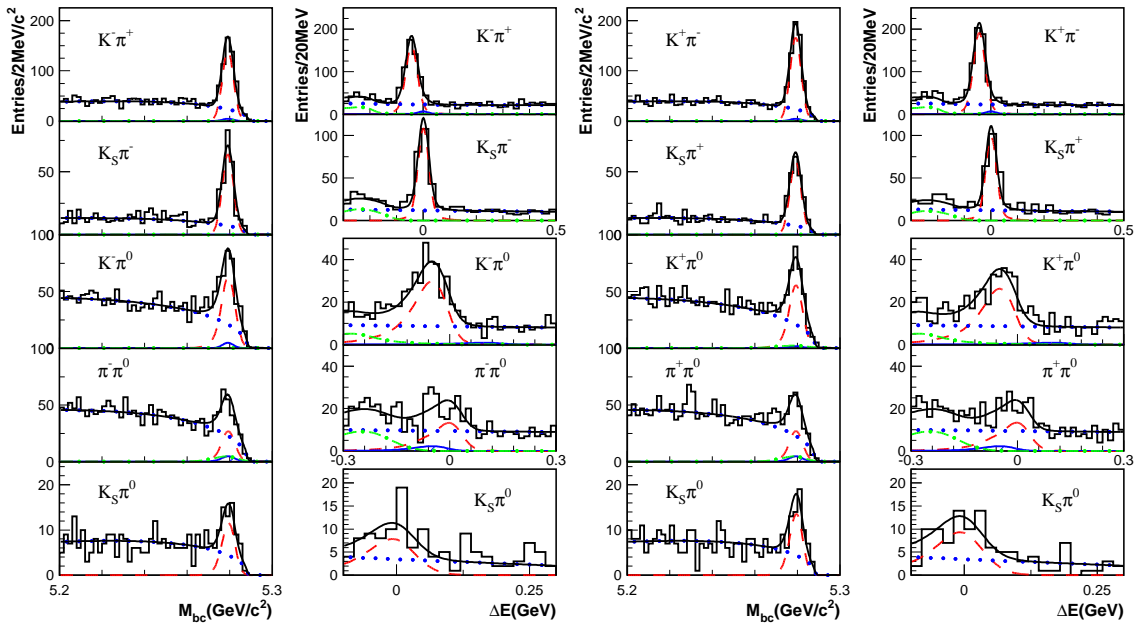,width=6.5in} 
\caption{$M_{\rm bc}$ (left) and $\Delta E$ (right) distributions for
$\overline B$ (left two columns) and $B$ (right two columns)
candidates reconstructed in the $K^\mp\pi^\pm$, $K^0_S\pi^\mp$, $K^\mp\pi^0$,
$\pi^\mp\pi^0$ and $K^0_S \pi^0$ modes (from top to bottom). The histograms
represent the data, while the curves represent the various components from
the fit: signal (dashed), continuum (dotted), three-body B decays
(dash-dotted), background from  mis-identification (hatched),
and sum of all components (solid).}
\label{fig:mbde_b0}
\end{figure*}

Table \ref{tab:acp} gives the signal yields and $\acp$ values for each mode.
The asymmetries for the background components are consistent with zero.
The errors on $\acp$ are determined from the ML fit and are consistent
with MC expectations.
Projections of the fits are shown in Fig.~\ref{fig:mbde_b0}.
The systematic errors from fitting are estimated by 
checking the deviations of the $\acp$ after varying each parameter of the 
signal PDFs by one standard deviation. The uncertainty in modeling 
the three-body background is studied by excluding the low $\Delta E$ region
($<-0.12$ GeV) and repeating the fit. For the $K^+ \pi^0$ and $\pi^+\pi^0$
modes, the uncertainty on the reflection-to-signal ratios obtained from 
$D^*$ control samples are included in the systematic error, with
contributions of $\pm0.006$ and $\pm0.009$ for the $K^+ \pi^0$ and $\pi^+\pi^0$
modes, respectively. The uncertainties in the mixing parameter $\chi_d$ and
wrong tag fraction $w_r$ are included in the systematic error for the
$K^0_S\pi^0$ mode. At each step, the deviation in $\acp$ is added in
quadrature to provide the systematic errors due to fitting,
which are  $^{+0.014}_{-0.020}$ for the $K^0_S\pi^0$ mode 
and less than 0.01 for all other modes. A possible detector bias in
$\acp$ is investigated using the
$B^+\to \overline{D}{}^0\pi^+$ samples, assuming null charge asymmetries in the
production. In this control sample, the same continuum suppression and
fitting procedures are applied. To mimic the $B^0\to K_S^0\pi^0$ analysis,
the flavor of the $B$ meson is
determined using flavor tagging information from the associated $B$.
Results of these null asymmetry checks are shown in Table \ref{tab:nullacp}.
The final systematic errors are then obtained by quadratically summing the
errors from the null asymmetry tests and the fitting systematics.

\begin{table}
\begin{center}
\caption{The fitted signal yields and $\acp$ results for
 individual modes.}
\begin{tabular}{lrc}
\hline\hline
~Mode~ & ~~~Signal Yield & $\acp$ \\
\hline
~$K^\mp\pi^\pm$ & $1029.8\pm35.3$ & $-0.088\pm0.035\pm0.013$~ \\
~$K^0_S \pi^\mp$ & $429.8\pm23.0$ & $+0.05\pm0.05\pm0.01$ \\
~$K^\mp\pi^0$ & $516.5\pm29.4$ & $+0.06\pm0.06\pm0.02$ \\
~$\pi^\mp\pi^0$ & $235.6\pm23.4$ & $-0.00\pm0.10\pm0.02$ \\
~$K^0_S\pi^0$ & $96.6\pm11.8$ & $+0.16\pm0.29\pm0.05$ \\
\hline\hline
\end{tabular}
\label{tab:acp}
\end{center}
\end{table}

\begin{table}
\begin{center}
\caption{Results of null asymmetry cross-checks.}
\begin{tabular}{lccc}
\hline\hline
Mode & $B \to D(\to K \pi) \pi$ & $B \to D(\to K \pi \pi^0) \pi$ &
 $B \to D\pi$ tagged\\
\hline
$N_{sig}$ & $6290.2\pm79.8$ & $7982.1\pm101.8$ & $8726.1\pm86.4$ \\
\acp & $0.002\pm0.013$ & $0.012\pm0.012$ & $-0.027\pm0.029$ \\
\hline\hline
\end{tabular}
\label{tab:nullacp}
\end{center}
\end{table}

In this study, the partial rate asymmetry $\acp(K^-\pi^+)$ is found to be
$-0.088\pm 0.035\pm 0.013$, which is $2.4\sigma$ from zero. The corresponding
90\% confidence level (C.L.) interval is $-0.15 < \acp(K^-\pi^+) < -0.03$.
Our central value is similar to that reported by BaBar,
$\acp(K^-\pi^+) = -0.107 \pm 0.041 \pm 0.013$ \cite{babar_pre},
indicating that the partial rate asymmetry may be negative. 
Theoretical predictions from different approaches suggest that
$\acp(K^-\pi^+)$ and $\acp(K^+\pi^0)$ should have the same sign.
The uncertainty in our result for $\acp(K^+\pi^0)$, is large
enough for it to be consistent with this expectation.
We set a 90\% C.L. interval of $-0.04 < \acp(K^+ \pi^0) < 0.16$.
Since no evidence of direct $CP$ violation is observed in
the decays $B^- \to K^0_S\pi^-$ and $B^+ \to \pi^+\pi^0$,
we set 90\% C.L. intervals: $-0.04 < \acp(K^0_S\pi^-) < 0.13$,
$-0.17 < \acp(\pi^+\pi^0) < 0.16$ and $-0.33 < \acp(K^0_S\pi^0) < 0.64$.

We thank the KEKB group for the excellent
operation of the accelerator, the KEK Cryogenics
group for the efficient operation of the solenoid,
and the KEK computer group and the NII for valuable computing and
Super-SINET network support.  We acknowledge support from
MEXT and JSPS (Japan); ARC and DEST (Australia); NSFC (contract
No.~10175071, China); DST (India); the BK21 program of MOEHRD and the
CHEP SRC program of KOSEF (Korea); KBN (contract No.~2P03B 01324,
Poland); MIST (Russia); MESS (Slovenia); NSC and MOE (Taiwan); and DOE
(USA). 


\end{document}